\documentclass[twocolumn,twoside,slac_two]{revtex4}
\usepackage{graphicx}
\usepackage{fancyhdr}
\newcommand{\rpv}{\mbox{$\not \hspace{-0.10cm} R_p \!$ }}
\def\lv{$\; \not \!\! L$}
\def\bv{$\; \not \!\! B$}
\def\l{\mbox{$ \lambda $}}
\def\lp{\mbox{$ \lambda' $}}
\def\lpp{\mbox{$ \lambda''$}}
\def\be{\begin{equation}}
\def\ee{\end{equation}}
\def\bea{\begin{eqnarray}}
\def\eea{\end{eqnarray}}

\def\greaterthansquiggle{\raise.3ex\hbox{$>$\kern-.75em\lower1ex\hbox{$\sim$}}}
\def\lessthansquiggle{\raise.3ex\hbox{$<$\kern-.75em\lower1ex\hbox{$\sim$}}}
\def\gl{\raise.3ex\hbox{$<$\kern-.68em\lower1ex\hbox{$>$}}}

\def\cv{{\cal V}}
\def\ll{\mbox{$l_L$}}
\def\er{\mbox{$e_R$}}
\def\ql{\mbox{$q_L$}}
\def\ur{\mbox{$u_R$}}
\def\dr{\mbox{$d_R$}}

\begin{document}

\title{Fermion Electric Dipole Moments in R-parity violating
Supersymmetry. }

%

\author{Rohini M. Godbole}
\affiliation{Centre for High Energy Physics, Indian Institute of Science, 
Bangalore, 560012, India.}

\begin{abstract}
In this talk I discuss some aspects of the study of electric dipole moments
(EDMs) of the fermions, in the context of R-parity violating (\rpv )
Supersymmetry (SUSY).  I will start with  a brief general discussion of 
how dipole moments, in general, serve as a probe of physics beyond the 
Standard Model (SM) and an even briefer summary of
\rpv SUSY. I will follow by  discussing  a general method of
analysis  for obtaining the leading fermion mass dependence of 
the dipole moments and present its application to \rpv\ SUSY case.
Then I will summarise the constraints that the analysis of $e,n$ and $Hg$ EDMs
provide for the case of trilinear \rpv SUSY couplings and make a few comments
on the case of bilinear \rpv, where the general method of analysis proposed
by us does not work.
\end{abstract}

\maketitle

\thispagestyle{fancy}
\section{\label{intro}Introduction}
\subsection{Fermion Dipole Moments}
Dipole moments of fermions in general, whether magnetic or electric and 
diagonal or transition, provide a very interesting probe of physics at the
loop level, since different invariance principles make very precise tree level
predictions for them. Any deviation from these then can give information
about loop-physics. Indeed electric dipole moments of the electron and 
neutron~\cite{electron,neutron} are an excellent probe of sources of
CP-violation beyond that available in the SM described by the CKM 
parametrisation.  The incredibly accurate test of Quantum Electro Dynamics 
(QED) provided by the measurement of $(g-2)_e$, possible signals for physics
beyond the Standard Model (SM) implied by the accurate measurement of 
$(g-2)_\mu$~\cite{Belanger:2001am} or constraints on all lepton number 
violating BSM physics from the lepton and flavour violating transition 
moments~\cite{Masiero:2004da} as well as  from the Majorana $\nu$ 
masses~\cite{Godbole:1992fb}, all show the very important role that the dipole
moments of fermions, in general, play in probing the loop level effects of
BSM physics that may be CP and flavour violating.

In the SM the CP odd neutron electric dipole moment (edm) $d_n$ vanishes at
two loops \cite{shabalin}. At three loops it has been
estimated~\cite{nshab,neutron} to be $d_n \sim 10^{-32 \pm
1}$~e~cm. Since there are no purely leptonic sources of CP violation in
the SM, an electron dipole moment can only be induced from $d_n$ at
second order in $G_F$ and thus may be estimated to be $d_e \sim (G_F
m_n^2)^2 d_n \sim 10^{-42} $~e~cm, to be compared to the estimate
$8\times 10^{-41}$~e~cm quoted in the literature \cite{booth, electron}.  
A nonzero $d_n$  can also induce an edm for $Hg$ or Deuteron.
The current experimental limits on the diagonal, fermion EDMs are $\sim
10^{-25}$ -- $10^{-26}$. For example, for the neutron, 
$d_n < 6.3 \times 10^{-25}$  e cm~\cite{PDG06}. Given the much 
lower estimates for
the same in the SM, it is clear that the fermion EDMs are a very promising 
place to look for footprints of CP violation in physics beyond the SM:
BSM physics. 
The experimental situation for the Deuteron and Mercury EDMs is more
competitive with respect to the theoretical predictions in the SM. 
Hence, these can be a test of the SM and can also put nontrivial constraints
on BSM physics.
Further,  with  additional sources of CP-violation,
beyond the one present in the SM,  EDMs for fermions ($e$ or $n$) may arise
even at the one-loop or two-loop level and thus provide strong constrains for 
the additional CP-violating  phases and/or new particle masses in the 
particular 
version of BSM. Supersymmetric theories ~\cite{susybooks} with or 
without R--parity violation~\cite{rpvint,Barbier:2004ez}, lepto-quark models, 
extended Higgs sectors are 
examples  of different types of BSM physics wherein additional sources of CP 
violation, which may or may not be flavour conserving, are strongly 
constrained by  considerations of the EDMs of fermions and neutral atoms 
like $Hg$~\cite{YaserAyazi:2006zw,Faessler:2006at}.  

\begin{figure}
  \includegraphics*[scale=0.40]{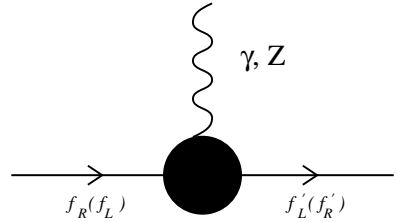}
 \caption{\label{fig:generic}Generic diagram which will contribute 
    to the dipole moment.}
 \end{figure}

Figure~\ref{fig:generic} shows a generic diagram which will contribute 
to the fermion dipole moment. One calculates in 
any  theory the matrix element of the current (electromagnetic 
or weak) as $q^\mu \to 0$. The EDMs, magnetic and electric, are then the
tensor form factors $F_T^\cv, F_T^{'\cv}$,  given by:
$$\bar u_{f_1} (p-q)\sigma_{\mu \nu} q^\nu (F_T^\cv + \gamma_5
F^{'\cv}_T)u_{f_2}(p), \ (\cv=\gamma, Z)$$

Dipole moment operators flip chirality, and hence have either to be
proportional to {\it some} fermion mass (this may not be the mass of the
external fermion), or to a chirality flipping Yukawa type 
coupling~\cite{barrzee}. 
The theoretical predictions for the moments of the heavier fermions like
the $t$, $b$ or the $\tau$, are larger than those for the first generation
particles due to the linear dependence on
$m_f$~\cite{hollik}. In models with lepto-quarks, particularly
large enhancements of the predicted values of the $\tau$ moments, by a
factor of $m_t/m_\tau$, are possible~\cite{taulq} and thus can be tested in
collider experiments. In this talk I will discuss a method of
analysis~\cite{Godbole:1999ye}, which allows to  extract the leading 
fermion mass
dependence of the coefficient of the induced dipole moment in any
theory.  The method is illustrated using the example of 
\rpv \cite{rpvint,Barbier:2004ez} SUSY  interactions.

\subsection{\rpv\ Supersymmetric Theories}
Supersymmetry, arguably, is the most attractive option for physics
beyond the SM. In these theories there exists  a discrete 
symmetry $R_p$ ($R_p = (-1)^{3B+L+2S}$) such that all the SM particles are 
even under it and all the superpartners are odd.  The $B,L$ and $S$ in the 
definition of $R_p$ above,  are the Baryon number, Lepton number and the 
spin of the particle respectively. Neither the conservation
nor the violation of $R_p$, is mandatory from a theoretical viewpoint. 
$B,L$ are symmetries of the SM but NOT of the MSSM. Hence the completely
general superpotential obtained by the requirement of gauge invariance and 
supersymmetry (SUSY),  contains the \rpv terms given by,
\bea
W_{{\not{R}}_p} = 
&\frac{1}{2}& \lambda_{ijk} L_i L_j E^c_k +
            \lambda'_{ijk} L_i Q_j D^c_k + \nonumber \\
&\frac{1}{2}& \lambda''_{ijk} U^c_i D^c_j D^c_k  +
            \kappa_i L_i H_2 .
\label{eq:eq1}
\eea
Here,  $L_i$, $Q_i$ are the doublet Lepton and Quark superfields,
$E_i$, $U_i, D_i$ are the singlet Lepton and Quark superfields. In general,
this superpotential contains terms violating  Baryon number (e.g.~$\lambda''$ 
terms) as well as the Lepton number (e.g.~$\l, \lp$). For
supersymmetric particles  with masses around a TeV, the former can cause very
rapid proton decay. This can be, however, cured by adopting $\lambda'' = 0$, 
corresponding to B conservation.  This choice is also preferred if we do not 
want the \rpv\ terms to wash out Baryon asymmetry generated through EW 
Baryogenesis. As a matter of fact, unified string theories actually prefer 
models with $B$ conservation and $R_p$ violation~\cite{Ibanez:1991pr}, as the 
former also suppresses the dimension five proton decay  operators. In addition,
non-zero $\nu$ masses can be generated in an economical way without 
introducing any new fields, only the \rpv couplings are 'new'. The bilinear
$\kappa_i's$ can generate the masses  at tree level whereas the trilinear
$\lambda,\lambda'$ terms generate them through quantum effects 
at one or two loop level~\cite{Borzumati:2002bf}. Superkamiokande and SNO 
have provided us with an
unambiguous proof of $\nu$ masses. In the \rpv\ Supersymmetric theories, there
exists enough freedom to generate the mass patterns indicated  by all the data.
Furthermore it leads to testable predictions at the colliders. All this makes 
\rpv SUSY worth detailed investigations. \rpv interactions can give rise to 
both the CP-conserving and CP-violating dipole moments of fermions; the latter
of course only when the couplings have non zero phases.

All the above positive things notwithstanding, one has to contend with the 
fact that allowing for this most general, \rpv\ superpotential, gives us a 
large number (total $48$) Yukawa type couplings, with NO theoretical 
indications about their sizes. On the positive note, many of these 
unknown couplings are constrained by a host of low energy 
processes~\cite{Barbier:2004ez,susybooks,Faessler:2006at}, such as the 
nucleon decay, $N$--$\bar N$ oscillations, $\mu$ decay,$B$--
physics,collider searches and last but not the least the fermion EDMs: 
the topic of present discussion.
For the $L$-- violating coupling the severest of the constraints come from 
$\nu$ masses~\cite{Borzumati:2002bf}.  In addition to this, 
serious constraints also exist from cosmology from considerations of 
Baryogenesis.  The EDMs provide some of the 
strongest constraints on the CP phases. 

In the next sections I would like to describe our work where we had developed a 
general method of analysis to obtain the leading fermion mass dependence of 
dipole moment operators in general, and of the EDMs, in particular. As 
mentioned before, large mass enhancements of the dipole moments were
observed in the lepto-quark models~\cite{taulq}.  Since in \rpv\ models the 
sfermions behave like lepto-quarks, it was interesting to check whether similar
enhancements obtain for \rpv SUSY as well. 

\section{Formalism and application to $d_e$ in \rpv\ SUSY.}
\subsection{Formalism}
Let us define five global charges $Q_{\ll},Q_{\er},Q_{\ql},
Q_{\dr}$ and $Q_{\ur}$ corresponding to five different $U(1)$ 
transformations, taking following values:
\bea
Q_{\ll} &=& 1\;\; {\rm for} \;\; e_{iL}, \nu_i,\widetilde e_{iL},
\widetilde \nu_i \;\;(i=1-3) \nonumber\\
&=& 0\;\; {\mbox{\rm for all the other (s)particles}}.
\label{eq:eq2}
\eea
The charge is independent of the generation. The value of all the charges
for all the Gauge bosons and Higgs bosons (plus their SUSY partners) are zero.
Note that the superpotential Yukawa interactions, $A$ terms and \rpv terms
do not conserve these charges where as the gauge (and gaugino) interactions do.
These charges are some kind of 'superchirality' in that they are nonzero
even for spin zero sfermions. They differentiate fermions of different 
chirality, and also right handed quarks of different electrical charge. They 
do not, however, differentiate between flavours of leptons or quarks with 
the same chirality, and so are conserved by inter-generational quark mixing.

Having introduced these charges, now we are ready to derive selection rules
that must be obeyed, for example, for the dipole moment operators. As mentioned
already and shown in Figure~\ref{fig:generic}, these operators require an 
interaction which will give rise to a flip
in chirality of the fermion. In the SM of course it is only the Yukawa
interaction terms. Once we go to the MSSM we also have the Higgsino mass
term as well as the trilinear $A$ term, which are proportional to the
same Yukawa coupling. In the MSSM the gaugino masses can also flip the 
chirality of gauginos, but in order for the chirality-flipped gaugino 
component to couple, one also needs $\tilde{f}_L-\tilde{f}_R$ mixing which 
has its origin in Yukawa interactions. In addition to these, in \rpv\ 
SUSY, there are also the \rpv interactions which can flip the fermion 
chirality.  We will call all these interactions 'Yukawa' interactions in
a generalised sense in what follows. 

For (say) a leptonic dipole moment 
to be generated we need 
$Q_{\ll}$ and $Q_{\er}$ to change by one unit in equal and opposite 
directions, with no change in the other charges. Similar change in  
$Q_{\ql}$ and $Q_{\ur}$ ($Q_{\dr}$ and $Q_{\ql}$) is required
for $u(d)$ dipole moments.  To be specific for the case of leptonic moments,
where $f,f'$ of Figure~\ref{fig:generic} are leptons, we must have:
\be
\Delta Q_{\ll} = -1, \Delta Q_{\er} = 1.
\label{eq:eq3}
\ee
or vice versa and all the other charges uncharged, i.e.
\be
\Delta Q_{\ql} = 0, \Delta Q_{\ur} = 0, \Delta Q_{\dr} = 0.
\label{eq:eq4}
\ee
Similar equations will hold for the case of the $u,d$ moments as well. 
Then, knowing the change induced in each of these 
charges by any (chirality-flipping) interaction,  it is
straightforward to derive relations between the number of vertices of
various types of chirality flipping interactions in order that these
collectively induce a dipole moment for any particular matter
fermion. 
Of course it clear that these would only be necessary conditions 
since, without further study, it cannot be guaranteed that the answer would 
not vanish. All this tells us is that it {\it need} not {\it vanish}.
Solving these conditions one can  then estimate the expected mass
dependence of the contribution of the diagram to the dipole moment operator.
Note also that our  method does not distinguish between the direct and 
transitional dipole moments.  So the results we obtain will be equally 
applicable to the case of (say) EDMs (in case of nonzero phases of the BSM 
couplings), as for the $\nu$ Majorana mass as well as the lepton number
violating decay like $\mu \rightarrow e \gamma$ etc.
\begin{table*}
\begin{center}
\caption{The change in the charges $Q_{\ll}, Q_{\er}, Q_{\ql},Q_{\ur}$
and $Q_{\dr}$ as defined in the text for different interactions that might
be present in SUSY models with MSSM field content. Gauge and gaugino
interactions or Higgs and Higgsino self interactions do not change any
of these charges.${\cal H}^0$ indicates any of the neutral Higgs bosons in the MSSM.}
\bigskip
\begin{tabular}{|c|ccccc|}
\hline
&&&&&\\
Interaction & $\Delta Q_{\ll}$ & $\Delta Q_{\er}$ &$\Delta Q_{\ql}$  &
$\Delta Q_{\ur}$ & $\Delta Q_{\dr}$ \\
&&&&&\\
\hline
Lepton Yukawa Interactions & -1 & +1 & 0 & 0 & 0 \\
Up quark Yukawa Interactions & 0 & 0 & -1 & +1 & 0 \\
Down quark Yukawa Interactions & 0 & 0 & -1 & 0 & +1 \\
${\cal H}^0 H^- \tilde d^*_R \tilde u_R$, $ H^- \tilde d^*_R \tilde u_R$
& 0 & 0 & 0 & -1 & +1 \\
\hline
$\l_{ijk} L_i L_j E_k^c$ interactions & -2 & +1 & 0 & 0 & 0 \\
$\lp_{ijk} L_i Q_j D_k^c$ interactions & -1 & 0 & -1 & 0 & 1 \\
$\lpp_{ijk} U_i^c D_j^c D_k^c$ interactions & 0 & 0 & 0 & 1 & 2 \\
\hline
\end{tabular}
\label{tab:one}
\end{center}
\end{table*}

Table~\ref{tab:one} shows the changes in various charges caused by different
interaction terms, for the SUSY model with the MSSM field content, allowing 
for the possibility of \rpv .  
Out of the Higgs-squark interaction terms arising from the $D$--term and 
the $F$--term, the trilinear Higgs-squark-squark and the 
Higgs-Higgs-squark-squark interaction terms cause the charges to change in 
a manner different from the above mentioned Yukawa interaction terms, as 
can be seen from the table. 

Using these, total change in a given charge in terms of the number of 
vertices of a given type present in the diagram can be written down trivially.
Let  $P,S$ and $R$ be the number of down-quark, up-quark and
lepton Yukawa interactions,
and  $P^*,S^*,R^*$ the  number of insertions corresponding to the 
Hermitean conjugate ($h.c.$) of these interactions.  $N,M,L (N^*,M^*,L^*)$  
denote  the number of vertices corresponding to interactions proportional to 
$\l , \lp , \lpp$ of Eq.~\ref{eq:eq1} respectively and 
$T (T^*)$ denote the number of trilinear or
quartic scalar vertices corresponding to the interactions in the fourth
row of Table~\ref{tab:one}.

The net change in various charges are given by
\bea
\Delta Q_{\ll} & = & -2 \Delta N - \Delta M -\Delta R \nonumber \\
\Delta Q_{\er} & = & \Delta N + \Delta R \nonumber\\
\Delta Q_{\ql} & = & -\Delta M -\Delta P -\Delta S \nonumber \\
\Delta Q_{\dr} & = & 2\Delta L + \Delta P  + \Delta M + \Delta T \nonumber \\
\Delta Q_{\ur} & = & \Delta L + \Delta S - \Delta T  ,
\label{eq:eq5}
\eea
where $\Delta M$, is given by $\Delta M = M - M^*$, {\it etc}.

Now we can solve this general system of equations
for the special cases of the moments of (say) leptons, 
by demanding that Eqs.~\ref{eq:eq3} and ~\ref{eq:eq4} are satisfied.
This general analysis and some numerical results  for the case of $d_e, d_n$ 
for the trlinear \rpv case are presented in the next section. 

\subsection{Expected fermion mass dependencies of the dipole moments.}
\noindent 
\underline{Leptonic Moments}:
\newline 
Let us start with the case of lepton moments. For any diagram to give a 
nonzero contribution to the dipole moment of a lepton,  Eqs.~\ref{eq:eq3} 
and ~\ref{eq:eq4} need to be satisfied.  Using Eqs.~\ref{eq:eq5} we then get,
\bea
\Delta N & = & 1- \Delta R \nonumber \\
\Delta M & = & \Delta R -1 \nonumber \\
\Delta P & = & 1- \Delta R -\Delta T \nonumber \\
\Delta L &=&0, \Delta S= \Delta T.
\label{eq:eq6}
\eea
\noindent
It is clear that any dipole moment ${\cal D}_l$ that this diagram can
give rise to will be
$$
{\cal D}_{l} \propto m_{l_i}^{R+R^*} m_{d_j}^{P+P^*}m_{u_k}^{S+S^*}
(m_{u_l}m_{d_l})^{T+T^*}
$$
with an appropriate numbers of the large masses (at least $M_W$ or
$M_{SUSY}$ depending on the graph) coming from the loops in the
denominator to give the right dimension.  Here, $m_{l_i}, m_{u_k}$ and
$m_{d_j}$ denote {\it some} lepton, up type quark and down type quark
mass. 

Notice that in the SM or in SUSY in absence of \rpv interactions,
we have $\Delta L = \Delta M = \Delta N =0.$ This in turn means that
$\Delta R =1$. I.e.,   $R$ or $R^*$ must be non-zero and hence the dipole 
moment has to be proportional to {\it some} lepton mass.  Since there is
no lepton flavour violation in the SM or the MSSM,  the moment 
$\propto m_l$. Further, in presence of \rpv, i.e, with nontrivial values 
of $\Delta M, \Delta N$, nonzero lepton  moment is possible ONLY with
an insertion of down-type OR  lepton Yukawa insertion. With \rpv there is 
also lepton flavour violation and  one may get  an enhancement of the lepton
moment  {\bf relative to SM/MSSM} by $m_b/m_l$.
The last of  Eqs.~\ref{eq:eq6}
tells us further that up-type quark masses enter only as even powers so
that these can never be the sole source of the required chirality flip
for a lepton dipole moment.  Indeed these masses have to be {\it in
addition to} the lepton or down type mass as mentioned above, and so
will necessarily be accompanied by the same power of some high mass in
the denominator, and so will actually suppress the moment.

\noindent 
\underline{down-type quark moments}:

\noindent
For this case it is the  Eqs.~\ref{eq:eq9} analogous to the earlier 
Eqs.~\ref{eq:eq6}  that need to be satisfied.
\bea
\Delta M & = & 1- \Delta P - \Delta T \nonumber\\
\Delta N & = & \Delta P -1 + \Delta T \nonumber\\
\Delta R & = & 1- \Delta P  -\Delta T \nonumber \\
\Delta L &=& 0,\Delta S=\Delta T.
\label{eq:eq9}
\eea
As in the previous case, we can easily see that for the SM/MSSM, the dipole
moments will be proportional to {\it some} down type quark mass, as it  
would vanish in the absence of all down-type Yukawa couplings.
Again the \rpv\ contributions to the dipole
moments of down-type quarks are nonzero only if either $\Delta R$ or
$\Delta P$ are non-zero,
and thus are proportional either to a lepton mass or a down-type quark
mass. Thus again no big enhancement involving the large top quark mass 
is possible.

\noindent
\underline{up-type quark moments}:

\noindent 
Now the conditions for the edm to be nonzero are given by
\bea
\Delta R & = & -\Delta N \nonumber\\
\Delta P & = & \Delta N - \Delta T \nonumber\\
\Delta M & = & - \Delta N \nonumber \\
\Delta L &=& 0,\Delta S=1 + \Delta T.
\label{eq:eq12}
\eea
In this case a solution {\it without} an up-type Yukawa interaction is
not allowed as opposed to the earlier two cases where a solution was
allowed where a single power of quark (lepton) mass could appear for the
lepton (quark) moment. Further, the leading mass dependence of an up-quark
moment generated by \rpv\ interactions is necessarily an up-type
mass.This happens because neither the \l\ or the \lp\ interactions
involve a $\tilde u_R$ or $u_R$.

We also see that for contributions that
will involve only the \lpp\ part of the \rpv\ interactions, the dipole
moment for the down quark will thus be proportional to $m_{d_i}
m_{u_i}^{2n}$ as opposed to the up quark moment which will be proportional
to $m_{u_i} m_{d_i}^{2n}$ ($n=0,1,2...$). This is in agreement with the
result for the edm due to the \lpp\ couplings that was derived long
ago~\cite{Barbieri:1985ty}.

One can make a few  more general comments. We observe that it is not 
possible to get an enhancement of the dipole moments by a factor
of $m_t/m_f$ in \rpv theories similar to the case of theories with general
lepto-quarks~\cite{taulq}, even though the squarks/sleptons
do play the role of lepto-quarks which have \lv\ or  \bv\
interactions.  This is simply due to the fact that  as opposed to the
case of a general lepto-quark, in SUSY with \rpv\, the couplings of the 
sfermions are chiral as a result of the supersymmetry.
This  allowed the charge assignment made in Eqs.~\ref{eq:eq2} in the
first place.
The chiral nature of the couplings, therefore, forbids the enhancement of
dipole moments of the leptons and down-type quarks  as compared to the
expectations in the SM/MSSM, by a  factor of $m_t/m_l$ or $m_t/m_d$.
Let us also add here that the mass dependencies obtained by our analysis
match with the results in the literature whenever explicit computations
are available. Refs.~\cite{Barbieri:1985ty,Chang:1998uc} are examples
of such earlier computations.
\noindent

\noindent 
\underline{Numerical estimates for EDMs}

\noindent 
Of course nowhere in the analysis so far we specialised to the case
of the edm. For the edm to be nonzero,  the diagram should be complex.
Again, if we look at the various conditions given by Eqs.~\ref{eq:eq6}, 
\ref{eq:eq9} and \ref{eq:eq12}, we see that to the lowest order, the \rpv\ 
contribution to a dipole moment  needs $N = N^* = 1$ or $M = M^* = 1$. 
Our framework then tells us that the diagram $\propto |\lambda|^2$ or 
$\lambda'|^2$. Thus it is clear that one loop diagram thus can not contribute 
to  edm, with just the trilinear \rpv couplings.

This analysis assumed NO lepton number violation in the sneutrino masses.
In the presence of such a violation, Majorana type $\nu$ masses required by 
SUSY and a left-right mixing in the charged slepton masses, one loop diagrams 
shown
\begin{figure}
  \includegraphics*[scale=0.4]{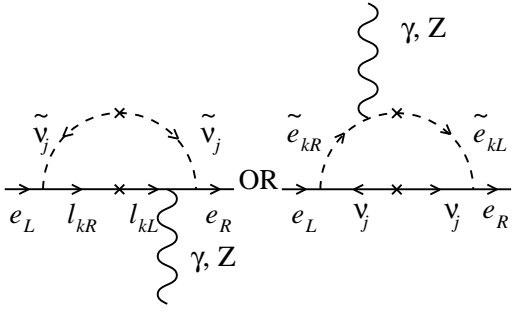}
 \caption{\label{fig:oneloop} Possible one loop diagrams
 diagram in \rpv theories,  which may contribute to the edm.}
 \end{figure}
in Figure~\ref{fig:oneloop} can exist and may give rise to an edm of the
electron. However, we expect such contribution to be severely limited by 
the constraints on the $\nu$ masses. For bilinear \rpv\ violation the
situation is different~\cite{Keum:2000ea} and I shall comment on it later.

Having established that the dominant contributions to the fermion EDMs from 
the trilinear \rpv couplings, can arise only at the two loop level, we 
studied various different types of two loop diagrams which would  achieve this.
Here I discuss a few examples.

\begin{figure}[htb]
\begin{center}
\includegraphics*[scale=0.4]{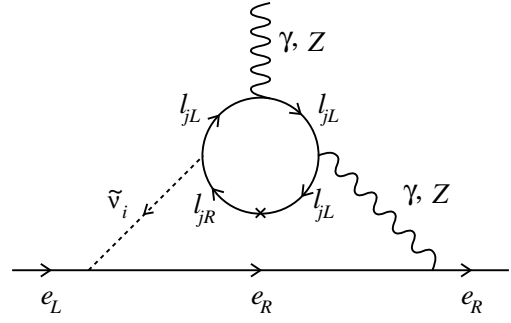}
\caption{An example of the  leading two-loop contribution to  the edm
of electron due to \l\  couplings.}
\label{fig:edml1}
\end{center}
\end{figure}
The diagram in Figure~\ref{fig:edml1}  corresponds to the case $N=N^*=1 $. 
As mentioned before, the contribution to the dipole moment then 
should be proportional to some $m_{l_j}$ which need not be the mass of
the external lepton (electron in this case).
This amplitude involves two {\it different} \l\ couplings and hence is
complex in general. Here the source of the complex nature of the
amplitude and hence for the edm, is the irremovable phases of the \rpv\
couplings.  The order of magnitude of the real part of the edm is estimated as 
the product of
explicit factors of couplings, mass insertions and colour factors, a
factor of $1/(4\pi^2)$ for each loop, and finally appropriate powers of
the ``large mass'' ($m_{\tilde \nu}$ in this case) in the denominator to
get the appropriate dimension. We then take the edm to be the imaginary
part ($\Im$) of this product. What we obtain is clearly an overestimate since
in practice the different diagrams may interfere destructively with each other.
For the diagram shown in Figure~\ref{fig:edml1} the contribution can be estimated 
to be:
\be
 d_e \sim {{(e^2,g_Z^2)} \over {4 \pi^2}} {1 \over {4 \pi^2}} \Im
\left[\sum_{ij,i\ne 1,j} m_{l_j} \lambda_{ijj}^{*} \lambda_{i11} {1
\over {m_{\tilde \nu _i}^2}}\right].
\label{eq:eq13}
\ee
Eq.~\ref{eq:eq13} reflects the enhancement by $m_\tau/m_e$ as expected from 
our earlier general analysis and it is noteworthy that this 
enhancement is obtained without paying any price for  mixing angles.

\begin{figure}[htb]
\begin{center}
\includegraphics*[scale=0.4]{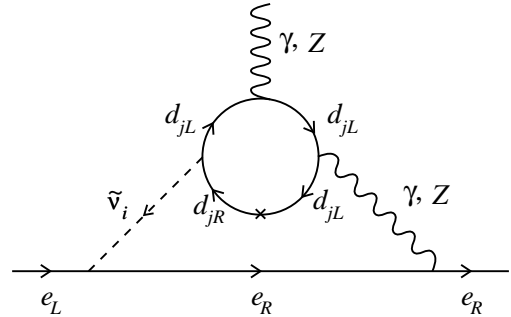}
\caption{An example of the  leading two-loop contribution to  the edm
of electron due to \l\ and \lp\  couplings.}
\label{fig:edml2}
\end{center}
\end{figure}
The diagram shown in Figure~\ref{fig:edml2} corresponds to $\Delta M
= -1, \Delta N= 1,\Delta P=1$ but with $\Delta R=0$. This  gives an
enhancement of the edm of the electron by a factor of ${m_b/m_e}$ as
we have already discussed and the dominant part of the
corresponding estimated contribution to $d_e$ is given by,
\be
d_e  \sim {{(e^2,g_Z^2)} \over {4 \pi^2}} {1 \over {4 \pi^2}}  m_b
\Im \left[\sum_{i\ne 1} 3 \lambda_{i33}^{'*} \lambda_{i11}
{1 \over {m_{\tilde \nu_i}^2}}\right],
\label{eq:eq14}
\ee
where a colour factor of 3 has been inserted.

Among the large number of possible two loop diagrams, those  
involving the Higgs exchanges will  all lead to contributions proportional to
$m_e$ or even higher powers in agreement with the expectations from
our general rules and others will be similar to the one given by
Eq.~\ref{eq:eq14} with the masses of the charged sleptons replacing those
of the sneutrino etc.

These two represent the dominant contributions and the current constraints
on the edm of the electron can be translated into a limit on products
of \l\ and \lp\ couplings, as given by Eq.~\ref{eq:eq15} below.
\bea \Im \left( \l_{211} \l_{233}^* \right ) &<& 5 \times
10^{-4} \left({{\tilde m}\over{1TeV}}\right)^2 \nonumber\\
\Im
\sum_{i \ne 1} \l_{i11} \lp_{i33}^*  & < & 0.6 \times 10^{-4}
\left({{\tilde m} \over {1 TeV}}\right)^2.
\label{eq:eq15}
\eea
Here $\tilde m$ stands for the mass of the appropriate SUSY
scalar. 

We can analyse the EDMs of the d-type quark and u-type quark in a similar 
fashion and then use those results in conjunction with the neutron edm
to constrain the different \rpv\ couplings. For example, diagrams for the $d$
quark similar to the one shown in Figure~\ref{fig:edml1}, replacing the
\l\ couplings by \lp\ couplings and the sparticle masses appropriately, lead
using  current experimental result $d_n < 6.0 \times 10^{-26} $~e~cm, to
\be
\Im \left [ \sum_{k} \lp_{k11} \lp_{k33}^* \right ] < 10^{-2}
\left({{\tilde m} \over {1 TeV}}\right)^2.
\label{eq:eq16}
\ee
Given the level of (justified) approximations made in getting our 
theoretical estimates for the quark edms, it did not make sense to 
include in our analysis the long distance effects in relating the neutron edm 
$d_n$ to $d_u,d_d$.  We further approximated $d_n$ by $d_d$~\cite{Godbole:1999ye}.

\subsection{Limits obtained using $d_{Hg}$}
As already mentioned in the introduction, for $d_{Hg}$ the SM predictions 
themselves can be competitive with the accuracy of the experimental 
measurements. A recent analysis of the hadronic edm in the presence of 
\rpv interactions~\cite{Faessler:2006at}, invokes one particular model
to relate the CP violation in the effective quark interactions to that 
giving rise to a hadronic edm.  They use a one pion-exchange model with CP-odd 
pion-nucleon interactions, generated through CP violating four quark 
interactions which in turn are caused  $\lambda'$ couplings. 
Limits on the \rpv couplings improve using the predictions for $d_{Hg}$ 
as compared to those from   $d_n$, by an order of magnitude.
Of course this limit has a model dependence.

\subsection{Bilinear \rpv}
As mentioned already, in the presence of bilinear \rpv, there are also soft 
SUSY breaking scalar bilinear terms  and our analysis  needs to be modified.
An important difference is that in the presence of the scalar
bilinear terms  the sneutrino fields generically acquire a $vev$, so that the
charge $Q_{l_L}$ is now no longer conserved. In principle, it would be
possible to include modifications to our analysis by allowing diagrams
where sneutrino fields disappear or are created from the vacuum: but the
result then depends on the number of fields that disappear into, or are
created from, the vacuum and the simple predictions that we have
obtained are lost. Our analysis clearly is inadequate 
when the bilinear mass term and the sneutrino VEV's are of the same order.
In fact  for the case of bilinear \rpv , dominant one loop \rpv contributions  
to the EDMs can exist~\cite{Keum:2000ea} and can usefully 
constrain, along with $\nu$ masses, the \rpv couplings. It may be interesting 
to do a more general analysis including both the bilinear and the trilinear 
\rpv interaction terms and understand the issue in an unified fashion.
\section{Conclusions}
In conclusion we presented a general analysis for obtaining the fermion mass
dependence of the induced dipole moment operator. 
We illustrated it with an example for the case of trilinear \rpv
interactions.
The estimates we obtain agree with the ones present in literature
when they are available. Our results show that the  unlike the lepto-quark
model big enhancements by factors of $m_t/m_f$ do not occur in this case.
The analysis needs to be modified for including the bilinear
\rpv terms.

\section{Acknowledgments}
It is a great pleasure to thank the organizers for this wonderful workshop
and school held in Tehran. Special thanks to Y. Frazan for inducing me to take
a fresh look at the recent developments in the subject.


\begin{thebibliography}{99} 
\bibitem{electron} For a review of the electron edm, see e.g., W.~Bernreuther
and M.~Suzuki, Rev. Mod. Phys. {\bf 63}, 313 (1991.
%
\bibitem{neutron} For a review of the neutron edm, see e.g., X-G.~He,
S.~Pakvasa and B.~McKellar, Int. J.~Mod. Phys. {\bf A4}, 5011 (1989).
%

\bibitem{Belanger:2001am}
See, for example, 
G.~Belanger, F.~Boudjema, A.~Cottrant, R.~M.~Godbole and A.~Semenov,
  Phys.\ Lett.\  B {\bf 519} (2001) 93
  [arXiv:hep-ph/0106275].

%
\bibitem{Masiero:2004da}
  A.~Masiero, S.~K.~Vempati and O.~Vives,
{\it  In *Tsukuba 2004, SUSY 2004* 627-636}
%
\bibitem{Godbole:1992fb}
 See e.g., R.~M.~Godbole, P.~Roy and X.~Tata,
  Nucl.\ Phys.\  B {\bf 401} (1993) 67
  [arXiv:hep-ph/9209251].
%
\bibitem{shabalin}  E.P.~Shabalin, Sov. J. Nucl. Phys. {\bf 28}, 75
(1978).
%
\bibitem{nshab}  E.P.~Shabalin, Usp. Fiz. Nauk. {\bf 139}, 561 (1983).
%
\bibitem{booth} I.~B.~Kriplovich and
M.~Pospelov,Sov. J. Nucl. Phys. {\bf B53}, 638 (1991); M.~J.~Booth,
hep-ph/9301293 find that the electron edm vanishes at three loops. F.~Hoogeven,
Nucl. Phys. {\bf B341}, 322 (1990), in an earlier computation, had
claimed a non-vanishing value
$d_e \sim 2 \times 10^{-38}$e~cm. at three loops.

\bibitem{PDG06}
 W.-M.Yao et al. (Particle Data Group), J. Phys. G 33, 1 (2006).


%
\bibitem{susybooks}
For a description of SUSY models, see
for example,
  M.~Drees, R.~M.~Godbole and P.~Roy,
  {\it Theory and Phenomenology of Sparticles},
  World Scientifc, Singapore (2005);
H. Baer and X. Tata, {\it Weak Scale Supersymmetry: 
From Superfields to Scattering Events}, Cambridge Univ. Press, (2006).

%
\bibitem{rpvint} L.~Hall and M.~Suzuki, Nucl. Phys. {\bf B231}, 419
(1984). 
\bibitem{Barbier:2004ez}
For a recent review see, for example, R.~Barbier {\it et al.}, 
  Phys.\ Rept.\  {\bf 420} (2005) 1,
  [arXiv:hep-ph/0406039].
%

\bibitem{YaserAyazi:2006zw}
  See, e.g.,  S.~Yaser Ayazi and Y.~Farzan,
  Phys.\ Rev.\  D {\bf 74} (2006) 055008
  [arXiv:hep-ph/0605272];
[arXiv:hep-ph/0702149], and references therein.
%
\bibitem{Faessler:2006at}
For some of the recent work on hadronic EDMs in the context of \rpv\ SUSY see,
e.g.,
  A.~Faessler, T.~Gutsche, S.~Kovalenko and V.~E.~Lyubovitskij,
  Phys.\ Rev.\  D {\bf 74} (2006) 074013
  [arXiv:hep-ph/0607269];
%
  Phys.\ Rev.\  D {\bf 73} (2006) 114023
  [arXiv:hep-ph/0604026];
  C.~C.~Chiou, O.~C.~W.~Kong and R.~D.~Vaidya,
  arXiv:hep-ph/0505207.


%
\bibitem{barrzee}S.~M.~Barr, E.~M.~Friere and A.~Zee,
Phys. Rev. Lett. {\bf 65}, 2626 (1990).
%
\bibitem{hollik} J.~Bernab\'eu, J. Vidal and G.A.~Gonz\'alez-Springberg,
Phys. Lett. {\bf B397}, 255 (1997); W. Hollik, J.~Ilana, S.~Rigolin and
D.~Stockinger, Phys. Lett. {\bf B425}, 322 (1998).
%
\bibitem{taulq} U. Mahanta, Phys. Rev. {\bf D54}, 3377 (1996);
W. Bernreuther, A.~Brandenburg and P.~Overmann,
Phys. Lett. {\bf B391}, 413 (1998); P. Poulose and S.~Rindani, Pramana
{\bf 51}, 387 (1998).
%
\bibitem{Godbole:1999ye}
  R.~M.~Godbole, S.~Pakvasa, S.~D.~Rindani and X.~Tata,
  Phys.\ Rev.\  D {\bf 61} (2000) 113003
  [arXiv:hep-ph/9912315].

%
\bibitem{Ibanez:1991pr}
  L.~E.~Ibanez and G.~G.~Ross,
  Nucl.\ Phys.\  B {\bf 368} (1992) 3.

%
\bibitem{Borzumati:2002bf}
  For a comprehensive analysis, of one and twoo-loop $\nu$ masses in \rpv\
theories see, e.g.,  F.~Borzumati and J.~S.~Lee,
  Phys.\ Rev.\  D {\bf 66} (2002) 115012
  [arXiv:hep-ph/0207184] and references therein.
%
\bibitem{Barbieri:1985ty}
  R.~Barbieri and A.~Masiero,
  Nucl.\ Phys.\  B {\bf 267} (1986) 679.
%
\bibitem{Chang:1998uc}
  D.~Chang, W.~Y.~Keung and A.~Pilaftsis,
  Phys.\ Rev.\ Lett.\  {\bf 82} (1999) 900
  [Erratum-ibid.\  {\bf 83} (1999) 3972]
  [arXiv:hep-ph/9811202].

\bibitem{Keum:2000ea}
  Y.~Y.~Keum and O.~C.~W.~Kong,
  Phys.\ Rev.\ Lett.\  {\bf 86} (2001) 393
  [arXiv:hep-ph/0004110];
  Phys.\ Rev.\  D {\bf 63} (2001) 113012
  [arXiv:hep-ph/0101113].

\end{thebibliography}
\end{document}